\documentclass[final,5p,times,authoryear]{elsarticle}

\usepackage{amssymb}
\usepackage{amsmath}
\usepackage{hhline}
\usepackage{color, colortbl}
\usepackage{booktabs}
\usepackage{multirow}
\usepackage{changepage}
\usepackage{pdflscape}
\usepackage{rotating}
\usepackage{hyperref}
\usepackage{nameref}
\usepackage{xcolor}
\usepackage{amsthm}
\usepackage{float}
\usepackage{caption}
\usepackage{graphics}
\usepackage[utf8]{inputenc}
\inputencoding{utf8}

\hypersetup{
    colorlinks,
    linkcolor={red!50!black},
    citecolor={blue!50!black},
    urlcolor={blue!80!black}
}

\journal{Computers $\&$ Operations Research}

\begin{document}
\begin{frontmatter}

\title{An efficient exact model for the cell formation problem with a variable number of production cells}
\author{Ilya Bychkov\corref{cor1}}
\ead{il.bychkov@gmail.com}

\author{Mikhail Batsyn\corref{cor2}}
\ead{mbatsyn@hse.ru}

\address{Laboratory of Algorithms and Technologies for Network Analysis,\\ National Research University Higher School of Economics,\\ 136 Rodionova, Nizhniy Novgorod, 603093, Russian Federation}

\cortext[cor1]{Corresponding author}

\begin{abstract}

The Cell Formation Problem has been studied as an optimization problem in manufacturing for more than 90 years.
It consists of grouping machines and parts into manufacturing cells in order to maximize loading of cells and minimize movement of parts from one cell to another.
Many heuristic algorithms have been proposed which are doing well even for large-sized instances.
However, only a few authors have aimed to develop exact methods and most of these methods have some major restrictions such as a fixed number of production cells for example. In this paper we suggest a new  mixed-integer linear programming model for solving the cell formation problem with a variable number of manufacturing cells. The popular grouping efficacy measure is used as an objective function. To deal with its fractional nature we apply the Dinkelbach approach. Our computational experiments are performed on two testsets: the first consists of 35 well-known instances from the literature and the second contains 32 instances less popular. We solve these instances using CPLEX software. Optimal solutions have been found for 63 of the 67 considered problem instances and several new solutions unknown before have been obtained. The computational times are greatly decreased comparing to the state-of-art approaches.

\end{abstract}


\end{frontmatter}

\newpage
\section{Introduction}
\label{sec_introduction}
The Cell Formation Problem as a part of Group Technology (GT) was introduced by \citet{5} and \citet{4}.
In the most general formulation it is designed to reduce production costs by grouping machines and parts into manufacturing cells (production shops).
The goal of such kind of grouping is to set up manufacturing process in a way that maximizes loading of machines within the cells and minimizes movement of parts from one cell to another.
In classical formulation the problem is defined by a binary matrix $A$ with $m$ rows representing machines and $p$ columns representing parts. In this machine-part matrix $a_{ij} = 1$ if part $j$ is processed on machine $i$.
The objective is to form production cells, which consist  of machines and parts together, optimizing some production metrics such as machine loading and intercell movement.

As an example of input data we will consider the instance of \citet{1} shown in Table~\ref{table1}. This instance consists of 5 machines and 7 parts.
The ones in a machine-part matrix are called \textit{operations}.
In Table~\ref{table2} a solution with 2 manufacturing cells is presented. The first manufacturing cell contains machines $m_{1}$, $m_{4}$ with parts $p_{1}$, $p_{7}$ and the second manufacturing cell contains machines $m_{2}$,$m_{3}$,$m_{5}$ with parts $p_{2}$,$p_{3}$,$p_{4}$,$p_{5}$,$p_{6}$. Some parts have to be moved from one cell to another for processing (e.g. part $p_{6}$ needs to be processed on machine $m_{1}$, so it should be transported from its cell 2 to cell 1).
The operations lying outside cells are called \textit{exceptional elements} or \textit{exceptions}. There can be also non-operation elements inside cells ($a_{ij}$ = 0). These elements reduce machine load and are called \textit{voids}.
So the goal is to minimize the number of exceptions and the number of voids at the same time.

\begin{table}[H]
\centering
        \begin{tabular}{ c|ccccccc|}
                  \multicolumn{1}{c}{~} & \multicolumn{1}{c}{$p_{1}$ } & \multicolumn{1}{c}{$p_{2}$ } & \multicolumn{1}{c}{$p_{3}$ } & \multicolumn{1}{c}{$p_{4}$ } & \multicolumn{1}{c}{$p_{5}$ } & \multicolumn{1}{c}{$p_{6}$ } & \multicolumn{1}{c}{$p_{7}$} \\ \hhline{~-------}
                  $m_{1}$ & \hspace{0.5em}1\hspace{0.5em}  & \hspace{0.5em}0\hspace{0.5em}  &\hspace{0.5em} 0\hspace{0.5em}  &\hspace{0.5em}0\hspace{0.5em}  &\hspace{0.5em}1\hspace{0.5em} &\hspace{0.5em}1\hspace{0.5em} &\hspace{0.5em}1\hspace{0.5em} \\ 
                  $m_{2}$ &0 &1 &1 &1 &1 &0 &0 \\ 
                  $m_{3}$ &0 &0 &1 &1 &1 &1 &0 \\ 
                  $m_{4}$ &1 &1 &1 &1 &0 &0 &0 \\ 
                  $m_{5}$ &0 &1 &0 &1 &1 &1 &0 \\ \hhline{~-------}
        \end{tabular}
        \caption{Machine-part 5 $\times$ 7 matrix from \citet{1}}
        \label{table1}
\end{table}

\begin{table}[H]
\centering
        \begin{tabular}{ c|ccccccc|}
                  \multicolumn{1}{c}{~} & \multicolumn{1}{c}{$p_{7}$ } & \multicolumn{1}{c}{$p_{1}$ } & \multicolumn{1}{c}{$p_{6}$ } & \multicolumn{1}{c}{$p_{5}$ } & \multicolumn{1}{c}{$p_{4}$ } & \multicolumn{1}{c}{$p_{3}$ } & \multicolumn{1}{c}{$p_{2}$} \\ \hhline{~-------}
                  $m_{1}$ &\hspace{0.5em}\cellcolor{yellow}1\hspace{0.5em}  & \cellcolor{yellow}\hspace{0.5em}1\hspace{0.5em}  &\hspace{0.5em}1\hspace{0.5em}  &\hspace{0.5em}1\hspace{0.5em}  &\hspace{0.5em}0\hspace{0.5em} &\hspace{0.5em}0\hspace{0.5em} &\hspace{0.5em}0\hspace{0.5em} \\ 
                  $m_{4}$ &\cellcolor{yellow}0 &\cellcolor{yellow}1 &0 &0 &1 &1 &1 \\ 
                  $m_{2}$ &0 &0 &\cellcolor{yellow}0 &\cellcolor{yellow}1 &\cellcolor{yellow}1 &\cellcolor{yellow}1 &\cellcolor{yellow}1 \\ 
                  $m_{3}$ &0 &0 &\cellcolor{yellow}1 &\cellcolor{yellow}1 &\cellcolor{yellow}1 &\cellcolor{yellow}1 &\cellcolor{yellow}0 \\ 
                  $m_{5}$ &0 &0 &\cellcolor{yellow}1 &\cellcolor{yellow}1 &\cellcolor{yellow}1 &\cellcolor{yellow}0 &\cellcolor{yellow}1 \\ \hhline{~-------}
        \end{tabular}
        \caption{Solution with 2 production cells}
        \label{table2}
\end{table}

\newpage
\subsection {Related work}
Many different approaches are proposed for solving the cell formation problem. The majority of them provide heuristic solutions and only a few exact methods have been suggested.

\citet{2} provided two MINpCUT exact models based on the well-known k-cut graph partition problem. The objective function considered in this research is minimization of the exceptional elements number for a fixed number of cells. Unfortunately this objective function does not address the load inside cells.

\citet{3} presented a mixed-integer linear programming model which maximizes the most popular objective for the cell formation problem - the grouping efficacy, introduced by ~\citet{6}. These authors suggested to apply Dinkelbach algorithm since the grouping efficacy is a fractional objective function. Their model allows solving the cell formation problem only if the number of production cells is predefined. Thus the suggested approach cannot guarantee global optimality of the obtained solutions with respect to a variable number of production cells. In many cases the computational times for this model are quite long or memory limitations are exceeded and the optimal solutions cannot be found.

\citet{7} introduced two approaches for solving the cell formation problem with the grouping efficacy objective. The first is a mixed-integer linear programming model which is based on a general two-mode clustering formulation with some simplifying assumptions (e.g. the numbers of clusters by rows and columns are equal). This model looks interesting, but requires too much time to be solved for many even medium-sized instances. The second approach is a branch-and-bound algorithm combined with a relocation heuristic to obtain an initial solution. The branch and bound approach is able to solve about two times more problem instances and the computational times are greatly improved as well. Generally it runs fine on well-structured (with grouping efficacy value $>$ 0.65 - 0.7) medium-sized problems.
Two major assumptions are made for both of these approaches: singletons are permitted (manufacturing cells containing only one machine or one part) that is quite a common practice;  residual cells are permitted (cells containing only machines without parts, or only parts without machines). Also the number of production cells is predefined for the both approaches, but for some test instances several values for the number of cells are considered in computational experiments.

Another model is provided in our earlier paper \citep{8}. There we present a mixed-integer linear programming formulation for the cell formation problem with a variable number of production cells. It deals well with small-sized instances,  but nevertheless the number of variables and constraints is huge -  O($m^2p$). This does not allow obtaining solutions even for some moderate-sized test instances and in some cases this model runs too slowly. 

Some authors used biclustering approaches to solve the cell formation problem.
\citet{64} applied simultaneous clustering for both dimensions (machines and parts) and minimized the number of voids plus the number of exceptional elements. \citet{62} reduced the cell formation problem to another biclustering problem - bicluster graph editing problem and suggested an exact method and a linear programming model which provides good computational results for the grouping efficacy objective.

\subsection {Contributions of this research}

In this paper we develop a fast compact model providing optimal solutions for the cell formation problem with a variable number of manufacturing cells and the grouping efficacy objective. Unlike the majority of linear programming models our model does not contain a direct assignment of machines or parts to cells. We use  machine-machine and part-machine assignments instead of the widely used machine-part-cell assignment. This leads to a compact and elegant formulation considering only constraints which ensure a block-diagonal structure of solutions. It allows us to drastically reduce the number of variables and constraints in our programming model and obtain globally optimal solutions even for some large-sized problem instances.

Computational experiments show that our model outperforms all known exact methods. We have solved 63 of 67 problem instances to the global optimum with respect to a variable number of production cells. We have also found several new solutions unknown before.

We would like to highlight that many researchers in the field use the 35 GT instances dataset provided by \citet{9}. These instances are taken from different cell formation research papers (references to the original sources are shown in Table \ref{testset_a_problems}). Some problem instances in this 35 GT dataset have errors and differ from the ones presented in the original papers. Many researchers including \citet{3} and \citet{62} have performed their computational experiments using these data from \citet{9}. We have reviewed all the original sources, comparing and forming the corrected version of this popular dataset. We have also collected many other problem instances less popular and formed a new dataset. All data can be downloaded from website \href{http://opt-hub.com/problems/cfp}{opt-hub.com} or \href{https://researchgate.net/publication/316648108_Test_instances_and_solutions_for_the_cell_formation_problem}{researchgate.net} (full urls can be found in references).

The paper is organized as follows. In Section \ref{sec_formulation} we provide the formulation of the cell formation problem.
Then in Section \ref{sec_model} we present our new mixed-integer linear programming model.
Sections \ref{sec_datasets} contains the information about datasetes we used for our experiments and  the computational results and comparisons to other exact approaches are given in Section \ref{sec_results}.
The conclusion is provided in Section \ref{sec_conclusion}.
\label {subsec_contrib}

\section{Problem formulation}
\label{sec_formulation}
Cellular manufacturing systems apply are aimed to process similar parts within the same production cell, balance machines workload and minimize parts movement from one cell to another during the production process. The most popular objective for the cell formation problem is the grouping efficacy introduced by \citet{6}:

\[
\tau =\frac{n_1^{in} }{n_1 +n_0^{in}
}
\] 
where

$n_1$ -- the total number of operations (ones) in the machine-part matrix,

$n_1^{in}$ -- the number of operations performed inside cells,

$n_0^{in}$ -- the number of voids (zeros inside cells). \newline

In comparisson to the other objectives the grouping efficacy function  addresses the best block-diagonal structure of the cell formation problem solutions \citep{63}.

In the literature several constraints related to the minimal size of a cell could be found. The following are the three most popular consideratons:

\begin{itemize}
  \item allowing residual cells (cells containing only machines or parts)
  \item allowing singletons (cells with one machine and several parts or vice versa) and prohibiting residual cells 
  \item allowing only cells with at least 2 machines and 2 parts \newline
\end{itemize}

The most popular option is allowing singletons and prohibiting residual cells. In this section for the classical formulation we assume that singletons can appear in solutions and residual cells are prohibited. In our new model and in computational experiments we consider the first two options. 

A straightforward integer fractional programming (IFP) model for the cell formation problem with the grouping efficacy objective function  allowing singletons and prohibiting residual cells is given below.
We use the following notation: $m$ is the number of machines, $p$ is the number of parts, $a_{ij}$ equals to 1 if machine $i$ processes part $j$ and $c$ is the maximal possible number of production cells. Since each production cell has to contain at least one machine and at least one part we set $c  = \min(m,p)$. \newline \\
\text{(IFP model):}\\

Decision variables: \newline
\[
 x_{ik} = 
  \begin{cases} 
   1,  & \text{if machine $i$ belongs to cell $k$,} \\
   0,        & \text{otherwise }
  \end{cases}
\] 
\[
 y_{jk} = 
  \begin{cases} 
   1, & \text{if part $j$ belongs to cell $k$, \quad} \\
   0,           & \text{otherwise }
  \end{cases}
\]\newline
\begin{equation}
\label{general_objective}
  max  \quad \frac{\sum_{i=1}^{m}\sum_{j=1}^{p}\sum_{k=1}^{c} a_{ij}x_{ik}y_{jk}}{\sum_{i=1}^{m}\sum_{j=1}^{p}a_{ij} +\sum_{i=1}^{m}\sum_{j=1}^{p}\sum_{k=1}^{c} (1 - a_{ij})x_{ik}y_{jk}}
\end{equation} 
\\

\text{Subject to:}
\\
\begin{equation}
\label{eq:machines_assigned}
\sum_{k=1}^{c} {x_{ik}}  = 1  \quad  i = 1,...,m
\end{equation} 

\begin{equation}
\label{eq:parts_assigned}
\sum_{k=1}^{c} {y_{jk}}  = 1  \quad j = 1,...,p
\end{equation} 

\begin{equation}
\label{eq:no_zerotones_1}
\sum_{i=1}^{m}x_{ik} \leq m\cdot\sum_{j=1}^{p}y_{jk}  \quad k = 1,...,c
\end{equation} 

\begin{equation}
\label{eq:no_zerotones_2}
\sum_{j=1}^{p}y_{jk} \leq p\cdot\sum_{i=1}^{m}x_{ik}  \quad k = 1,...,c
\end{equation} 

Objective function \eqref{general_objective} is the grouping efficacy measure where the numerator is the number of ones inside cells ($n_1^{in}$) and two sums in the denominator are the total number of ones ($n_1$) and the number of zeros inside cells ($n_0^{in}$) respectively.
Constraints \eqref{eq:machines_assigned} and \eqref{eq:parts_assigned} require that each machine and each part is assigned to exactly one production cell.
The following two inequalities \eqref{eq:no_zerotones_1} and  \eqref{eq:no_zerotones_2} prohibit residual cells (without machines or  parts). The left part of \eqref{eq:no_zerotones_1} is the total number of machines assigned to the particular cell (this sum is not greater than $m$) and the right part is the total number of parts assigned to that cell (multiplied by $m$). It means that if we have at least one machine assigned to some cell there should be at least one part assigned to this cell. This model allows us to have any number of cells in the optimal solution. For example if optimal solution has only two cells then variables 
$x_{ik}$ and $y_{jk}$ will be zero for all $k$ except only two values of $k$.

\section{MILP model}
\label{sec_model}
\subsection{Objective linearization}
In our paper ~\citet{8} we have proposed a mixed-integer linear programming model for the cell formation problem which is very similar to the one described in the previous section. One of the most important points there was linearization of the grouping efficacy objective. Our previous idea was to linearize the grouping efficacy objective function by fixing the value of denominator $n_1 +n_0^{in}$ and considering a range of all possible numbers of voids $n_0^{in}$. The lower bound for $n_0^{in}$ equals to 0 because generally there can be a solution without any voids. The upper bound is computed using the following proposition.

\newtheorem*{thm*}{Proposition 1 \citep{8}}
\begin{thm*}
The number of voids in the optimal solution satisfies the following inequality:
\[
n_0^{in} \leq \left\lfloor \frac{1 - \tau}{\tau}n_{1} \right\rfloor
\]
where $\tau$ is the grouping efficacy value of any feasible solution.
\end{thm*}

So if we know a feasible solution we can limit the range of possible values for the number of voids. Unfortunately, the performance of this approach strongly depends on the feasible solution we use for 
obtaining our bounds. This way solving problem instances where grouping efficacy value is low takes too much computational resources (since the number of subtasks is too large) and sometimes we are unable to solve even medium-sized cell formation instances.

In this paper together with using our new mixed-integer linear model we use another way of linearization -- \citet{65} algorithm.
The parametric approach introduced by W.Dinkelbach is one of the most general and popular strategies for fractional programming. It reduces the solution of a fractional programming problem to the solution of  a sequence of simpler problems.
If we consider a fractional programming model with the following objective function:

\begin{equation}
\label{fract-lin-objective}
Q(x) = \frac{P(x)}{D(x)},
\end{equation} 
then Dinkelbach procedure is the following: 

\begin{itemize}
  \item \textbf{Step 1} take some feasible solution $x^{0}$, compute $\lambda_{1} = \frac{P(x^{0})}{D(x^{0})}$ and  let $k=1$
  \item \textbf{Step 2} solve the original problem with objective function $Q(x)$ replaced with  $F(\lambda_{k}) =  P(x) - \lambda_{k}D(x) \rightarrow max$
and let $x^{k}$ be the optimal solution
  \item \textbf{Step 3} If $F(\lambda_{k})$ is equal to $0$ (or less than some predefined tolerance) then stop the procedure and return $x^{k}$ as the optimal solution. \newline
  Else $ k = k+1, \lambda_{k} = \frac{P(x^{k})}{D(x^{k})}$ and goto step 2.
\end{itemize}

 \citet{3} have also used Dinkelbach approach for linearization of grouping efficacy measure. Although their computational times are quite high and the results are given only for the particular fixed number of production cells.

\subsection{Suggested two-index model}
\label{subsec_2ind}
Due to a large number of variables and constraints in three-index model \citep{8} CPLEX spends too much computational resources solving even small-sized instances (we have solved only 14 of 35 problem instances).
Here we introduce a two-index mixed-integer linear programming model for the cell formation problem. The key idea of this model is removing machine-part-cell 
relation as it has been done in many works before. Instead of mapping elements to cells we use a simple fact that machines within the same production cell have the same set of parts assigned to that cell. The two-index model combines well with the Dinkelbach algorithm and shows impressing performance even on large-sized problem instances.

Two-index model:
\[
 x_{ik} = 
  \begin{cases} 
   1,  & \text{if machines $i$ and $k$ are in the same cell,\quad } \\
   0,        & \text{otherwise }
  \end{cases}
\] 
\[
 y_{ij} = 
  \begin{cases} 
   1, & \text{if machine $i$ and part $j$ are in the same cell,} \\
   0,           & \text{otherwise }
  \end{cases}
\]
\\
\begin{equation}
\label{2IND_obj}
max \quad\sum_{i=1}^{m} \sum_{j=1}^p a_{ij}y_{ij} - \lambda \cdot(\sum_{i=1}^{m} \sum_{j=1}^p (1-a_{ij})y_{ij} + \sum_{i=1}^{m} \sum_{j=1}^p a_{ij})
\end{equation} 
\text{Subject to:}
\\
\begin{equation}
\label{2IND_rect1}
2x_{ik} - y_{ij} - y_{kj} \geq -1 \quad  i,k = 1,\dots,m  \quad  j = 1,\dots,p 
\end{equation} 
\begin{equation}
\label{2IND_rect2}
y_{ij} - y_{kj} - x_{ik} \geq -1 \quad  i,k = 1,\dots,m  \quad  j = 1,\dots,p  
\end{equation}
\begin{equation}
\label{2IND_rect3}
y_{kj} - y_{ij} - x_{ij} \geq -1 \quad  i,k = 1,\dots,m  \quad  j = 1,\dots,p  
\end{equation}
\begin{equation}
\label{2IND_machine_with_part}
\sum_{j=1}^{p}y_{ij} \geq 1 \quad  i = 1,...,m
\end{equation} 
\begin{equation}
\label{2IND_part_with_machine}
\sum_{i=1}^{m}y_{ij} \geq 1 \quad  j = 1,...,p
\end{equation}

Technically matrix $[x_{ik}]$ here can be replaced by the one with part-part relations, however the number of machines in problem instances is usually lower than the number of parts (for large-sized instances the difference is significant). As a result we have $m^2$ variables from matrix $[x_{ik}]$ and $mp$ variables from matrix $[y_{ij}]$.

Objective function \eqref{2IND_obj} is the grouping efficacy measure linearized using Dinkelbach method. Constraints \eqref{2IND_rect1}, \eqref{2IND_rect2}, \eqref{2IND_rect3} set relations between machines and parts to ensure the solution can be transformed into the block-diagonal form (which means its feasibility). The last two inequalities \eqref{2IND_machine_with_part} and \eqref{2IND_part_with_machine} are optional and prohibit residual cells. 

We start with setting $\lambda$ equal to the best known efficacy value for the considered cell formation problem instance. Then we sequentially solve several two-index problems according to the Dinkelbach algorithm and update $\lambda$ value with the solutions found until our objective function is above 0.
\begin{table}[H]
  \centering
  \scriptsize
  \caption{Testset A - Instances}
    \resizebox{\columnwidth}{!}{%
    \begin{tabular}{cccc}
    \toprule
    \multirow{2}[2]{*}{\textbf{ID}} & \multirow{2}[2]{*}{\textbf{Source}} & \multirow{2}[2]{*}{\textbf{m}} & \multirow{2}[2]{*}{\textbf{p}} \\
         &      &      &  \\
   \midrule
    A1   & \citet{13} - Figure 1a& 5    & 7 \\
    A2   & \citet{14} - Problem 2 & 5    & 7 \\
    A3   & \citet{15} & 5    & 18 \\
    A4   & \citet{16} & 6    & 8 \\
    A5   & \citet{17} & 7    & 11 \\
    A6   & \citet{18} - Example 1 & 7    & 11 \\
    A7   & \citet{19} & 8    & 12 \\
    A8   & \citet{20} & 8    & 20 \\
    A9   & \citet{21} & 8    & 20 \\
    A10  & \citet{22} & 10   & 10 \\
    A11  & \citet{24} & 15   & 10 \\
    A12  & \citet{25} & 14   & 24 \\
    A13  &\citet{26} & 14   & 24 \\
    A14  & \citet{27} & 16   & 24 \\
    A15  & \citet{10} & 16   & 30 \\
    A16  &\citet{28} & 16   & 43 \\
    A17  & \citet{29} & 18   & 24 \\
    A18  & \citet{23} & 20   & 20 \\
    A19  & \citet{30} & 23   & 20 \\
    A20  & \citet{29} & 20   & 35 \\
    A21  & \citet{31} & 20   & 35 \\
    A22  & \citet{32} - Dataset 1& 24   & 40 \\
    A23  & \citet{32} - Dataset 2& 24   & 40 \\
    A24  & \citet{32} - Dataset 3& 24   & 40 \\
    A25  & \citet{32} - Dataset 5& 24   & 40 \\
    A26  & \citet{32} - Dataset 6& 24   & 40 \\
    A27  & \citet{32} - Dataset 7& 24   & 40 \\
    A28  & \citet{27} & 27   & 27 \\
    A29  & \citet{29} & 28   & 46 \\
    A30  &\citet{33} & 30   & 41 \\
    A31  & \citet{26} - Figure 5 & 30   & 50 \\
    A32  & \citet{26} - Figure 6 & 30   & 50 \\
    A33  &\citet{34} & 30   & 90 \\
    A34  & \citet{27} & 37   & 53 \\
    A35  & \citet{35} & 40   & 100 \\
    \bottomrule
    \end{tabular}%
    }
  \label{testset_a_problems}%
\end{table}%
\section{Test instances}
\label{sec_datasets}

For our computational experiments we have used two datasets, \textit{Testset A} and \textit{Testset B}.
  
\textbf {Testset A - Classic}. The first dataset is a classical 35 GT problem set taken from \citet{9}. It contains 35 test instances with
sizes from $5 \times 7$ up to $40 \times 100$ (machines $\times$ parts notation). This dataset is very popular among cell formation researchers and there are lots of computational results obtained by different methods (heuristics and metaheuristics generally). As we highlighted before some problem instances in this dataset have inconsistencies with the original papers they are published in. We have compared these instances to the original ones and corrected the dataset.

\textbf{Testset B - Extra}.
Another dataset named \textit{Testset B} consists of other instances taken from different papers. We have looked through many papers on the cell formation problem and formed this new set. There are 32 test instances with sizes from $6 \times 6$ to $50 \times 150$. A couple of instances from this set have been adopted to the classical formulation of the cell formation problem.

Since the number of machines determines the size of our model  (the number of decision variables and constraints) we consider 3 classes of problem instances.
\begin{itemize}
  \item small (less than 10 machines)
  \item medium (from 10 to 20 machines)
  \item large (20 machines or greater)
\end{itemize}

For our data we can conclude that Testset A has 2 times more large instances, but  less medium and small instances (see Table~\ref{instances_sizes}).

\begin{table}[htbp]
  \centering
  \normalsize
  \caption{Testsets instances size}
    \begin{tabular}{cccc}
    \toprule
         & small & medium & large \\
    \midrule
    Testset A & 9    & 8    & 18 \\
    Testset B & 11   & 13   & 8 \\
    \bottomrule
    \end{tabular}%
  \label{instances_sizes}%
\end{table}%

\section{Computational results}
\label{sec_results}
For our computational experiments we consider two most popular versions of cell size constraints:
\begin {enumerate}
\item Residual cells are prohibited, singletons are allowed (each cell has at least 1 machine and 1 part)
\item Residual cells are allowed (cells with only machines or only parts can appear in the final solution)
\end {enumerate}

Several state-of-art exact approaches have been chosen for comparisons. 
As a platform for our computations we have used  a system with Intel Xeon processor running at 3.4 GHz with 16GB RAM and CPLEX 12.4.0 installed.
Due to high-quality initial solutions the Dinkelbach algorithm makes only one or, in rare cases, two iterations.
\begin{table}[H]
  \scriptsize
  \caption{Testset B - Instances}
    \resizebox{\columnwidth}{!}{%
    \begin{tabular}{cccc}
    \toprule
    \multirow{2}[2]{*}{\textbf{ID}} & \multirow{2}[2]{*}{\textbf{Source}} & \multirow{2}[2]{*}{\textbf{m}} & \multirow{2}[2]{*}{\textbf{p}} \\
         &      &      &  \\
   \midrule
    B1    & \citet{39} & 6    & 6 \\
    B2    &\citet{40} & 6    & 7 \\
    B3    & \citet{41} & 6    & 11 \\
    B4   & \citet{24} & 7    & 5 \\
    B5   & \citet{17} & 7    & 8 \\
    B6   & \citet{42} & 7    & 8 \\
    B7   & \citet{44} & 7    & 10 \\
    B8   & \citet{45} & 8    & 8 \\
    B9  & \citet{46} & 8    & 10 \\
    B10   &\citet{47} & 8    & 10 \\
    B11   & \citet{30} & 9    & 15 \\
    B12   & \citet{49} & 10   & 8 \\
    B13   & \citet{50} & 10   & 12 \\
    B14   &\citet{43} & 10   & 38 \\
    B15   & \citet{44} & 11   & 10 \\
    B16   & \citet{51} & 11   & 22 \\
    B17   & \citet{53} & 12   & 19 \\
    B18   & \citet{37} & 14   & 14 \\
    B19   & \citet{24} - Fig.3a & 15   & 10 \\
    B20   & \citet{12} - Fig.6b & 15   & 15 \\
    B21   & \citet{12} - Fig.6c & 15   & 15 \\
    B22   & \citet{12} - Fig.6d & 15   & 15 \\
    B23   & \citet{54} & 17   & 20 \\
    B24   & \citet{55} & 18   & 24 \\
    B25   & \citet{58} & 20   & 10 \\
    B26   & \citet{59} & 20   & 51 \\
    B27   & \citet{44} & 26   & 28 \\
    B28   & \citet{12} - Fig.7 & 28   & 35 \\
    B29   & \citet{60} & 35   & 15 \\
    B30   & \citet{60} & 41   & 50 \\
    B31   & \citet{12} - Fig.12 & 46   & 105 \\
    B32   & \citet{61} & 50   & 150 \\
    \bottomrule
    \end{tabular}%
    }
  \label{testset_b_problems}%
\end{table}%

\subsection{Testset A}
\subsubsection{Experiments}
The instances from Table~\ref{testset_a_problems} have been studied widely in the literature.
We report results separately for the formulation where minimal cell size is $1 \times 1$ (Table~\ref{testset_a_single} and Figure~\ref{testset_a_single_running_times}) and the formulation with residual cells allowed (Table~\ref{testset_a_residual}  and Figure~\ref{testset_a_residual_running_times}).
In the first case we have selected two approaches for the results comparison:
\begin{enumerate}
  \item The  MILP model by \citet{3}
  \item  The MILP model by \citet{8}
\end{enumerate}
\citet{3} considered a simplified formulation of the cell formation problem solving every problem instance only for one fixed number of production cells. 
These authors have performed computational experiments on an AMD processor 2.2 GHz with 4GB RAM.
For Testset A we use the best efficacy results from the literature as initial values for $\lambda$ parameter.

In case of unrestricted cell sizes (residual cells are allowed) we have compared our results to the following approaches:
\begin{enumerate}
  \item The branch-and-bound algorithm by \citet{7}
  \item The ILP model by \citet{62}
  \item The iterative exact method by \citet{62}
\end{enumerate}

\citet{7} considers several values for the number of cells for some problem instances, so in this case we compare our computational time with these times summed up for every test instance. 
As hardware platforms \citet{7} reports 3.4 GHz Intel Core i7-2600 with 8GB RAM and \citet{62} the same 3.4 GHz Intel Core i7-2600  with 32 GB RAM.

Since \citet{3} and \citet{7} do not consider all possible numbers of production cells we show the number of cells in brackets for these approaches.

\subsubsection{Results}
The results for Testset A are summarized in Table~\ref{testset_a_single} and Table~\ref{testset_a_residual}. For each algorithm we report  the grouping efficacy value and the running time in seconds. Since our testset is larger than the one used by \citet{7} the missing results are marked as ''-''. For some problems exact solutions have not been obtained because CPLEX runs too long or takes too much memory. These instances are marked as "*". 
\begin{figure}[H]
    \centering
    \includegraphics[scale=0.35]{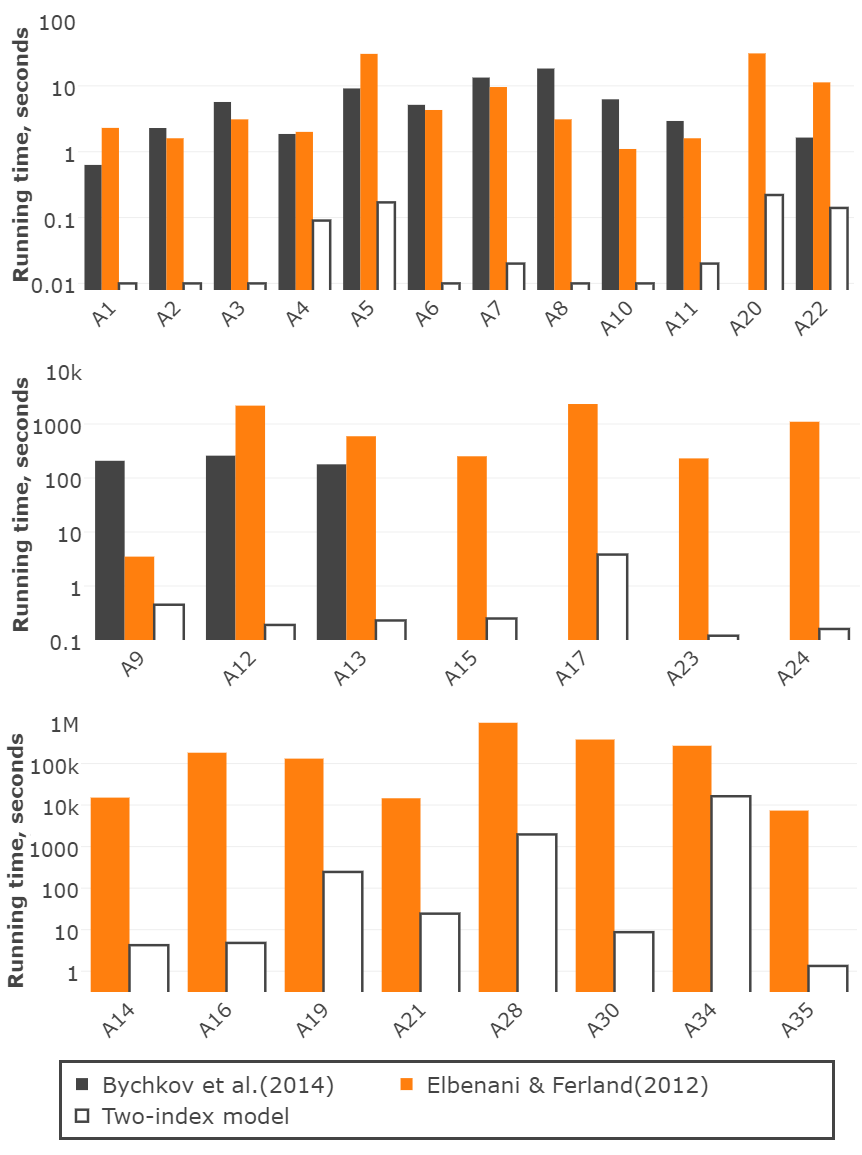}
    \caption{Testset A - No residual cells. Running times comparison.}
    \label{testset_a_single_running_times}
\end{figure}
Table~\ref{testset_a_single} shows the results for the case where we prohibit cells without machines or parts. Our previous model from \citet{8} also considers a variable number of production cells, but due to its complexity and not very effective linearization technique it is able to solve only 14 test problems of 35. The model suggested by \citet{3} solved 27 problem instances but only for the one fixed number of production cells for each problem instance. Our new model provides global optimal solutions (with respect to any possible number of cells) for 31 of 35 problem instances.  For problem instance A33 we have found a new solution with grouping efficacy 0.48 unknown before.

For 17 instances: A14-A21, A23-A26, A28, A30, A31, A34 and A35 we are the first to prove the global optimality of the best known solutions with respect to a variable number of production cells.

\begin{figure}[H]
    \centering
    \includegraphics[scale=0.35]{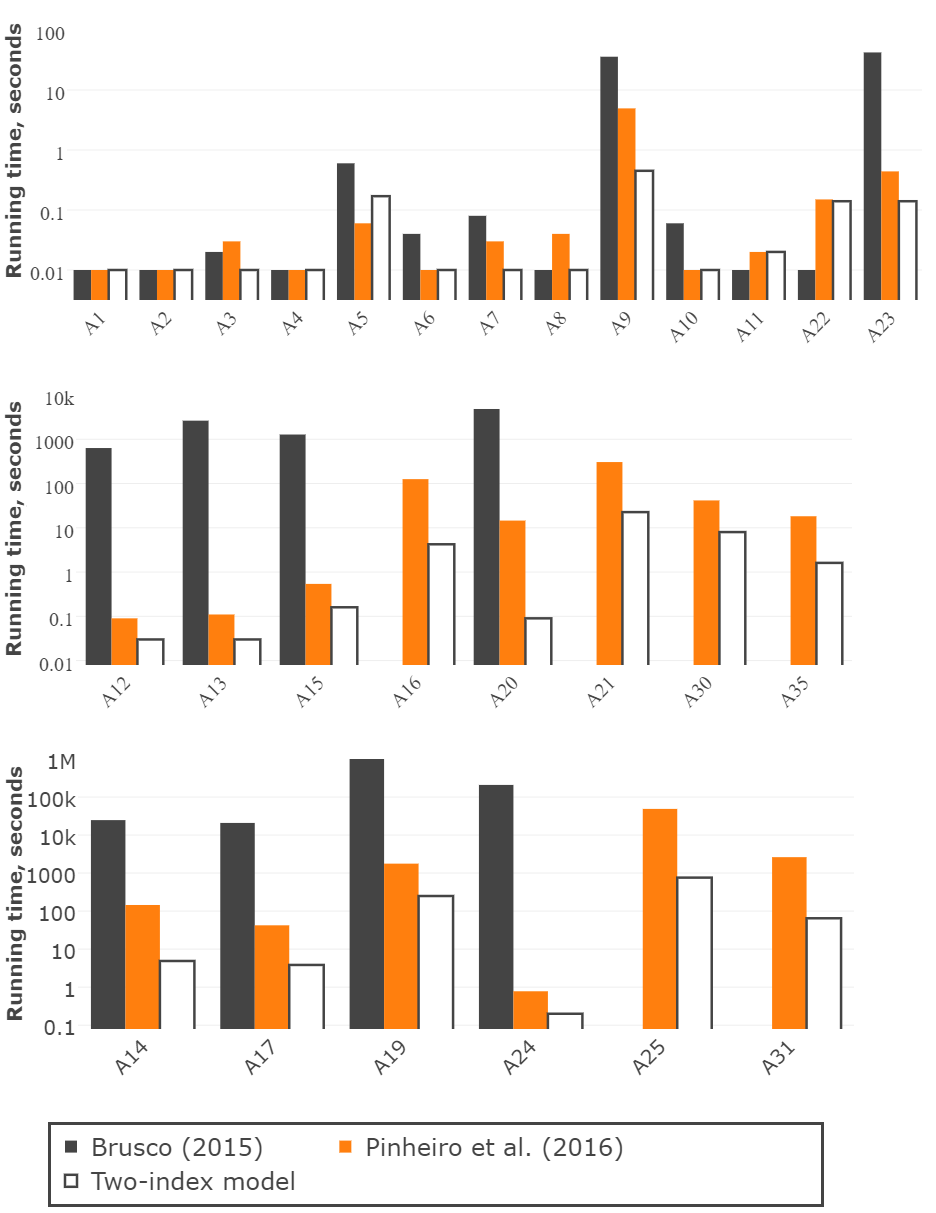}
    \caption{Testset A - Allowed residual cells. Running times comparison.}
    \label{testset_a_residual_running_times}
\end{figure}

Running times bar chars for Table~\ref{testset_a_single} are presented in Figure~\ref{testset_a_single_running_times}. Here we have used logarithmic scale with base 10 for Y axis (running time). Our new model shows really good performance, it works from 7 to 43383 times faster than the model from \citet{3} and from 11 to 1833 times faster than the model from \citet{8}.  We must underline that although we use a better hardware platform than \citet{3}, our problem formulation is more complicated than a formulation with a fixed number of cells.

The results for the formulation with no constraints on cell sizes are summarized in Table~\ref{testset_a_residual}. The model suggested by \citet{62} solved 27 problem instances to the global optimum. Our approach has obtained exact solutions for 32 of 35 test instances. In addition for problem instances A18, A33 and A34 we have found new solutions unknown before.

Running times bar charts for Table~\ref{testset_a_residual} are presented in Figure~\ref{testset_a_residual_running_times}. Here we have chosen the ILP model from \citet{62} for comparison since it has a better performance than the exact iterative method of the same authors. In Figure~\ref{testset_a_residual_running_times} we have also used logarithmic scale with base 10 for the first and second plots (instances with running times less than 60 seconds and less than 5000 seconds). For the last plot (instances with running times less than 1500000 seconds) we have used logarithmic scale with base 100. We can see that  the two-index model runs up to 1 million times faster than the branch-and-bound algorithm by \citet{7} and up to 161 times faster than the ILP  model by \citet{62}.

\subsubsection{Inconsistencies}
The classical dataset of 35 GT problems from \citet{9} have been used for many years by the cell formation researchers for computational experiments and results comparison.
Unfortunately, the dataset contains several inconsistencies with the original sources: papers from \citet{13} to \citet{35} (see Table~\ref{testset_a_problems}). Many researchers have used corrupted instances and sometimes add some new inconsistencies. Therefore obtaining results for these problems and comparing it to results of other approaches becomes a really difficult task. One of the goals of this paper is to provide correct data for the cell formation researchers. In this paper we mark usage of inconsistent data with superscript \textsuperscript{E}. 

We have not been able to obtain results reported in \citet{3} for problem instances A15 and A31 using both possible data sets - dataset from  \citet{9} and our corrected version. Probably some other data have been used.

\begin{table}[h]
 \centering
 \captionsetup{font=small}
 \setlength{\tabcolsep}{4pt}
  \small
  \caption{Computational experiments on the data provided by \citet{9} }
    \begin{tabular}{|c|rr|cc|}
    \toprule
    \multicolumn{1}{|c|}{\multirow{2}{*}{\textbf{\#}}}  &  \multicolumn{2}{c|}{\textbf{Time, sec}} & \multicolumn{2}{c|}{\textbf{Efficacy}} \\
    \multicolumn{1}{|c|}{} &   \multicolumn{1}{c|}{Pinheiro et al.} &  \multicolumn{1}{c|}{two-index} &  \multicolumn{1}{c|}{Pinheiro et al.} &  \multicolumn{1}{c|}{two-index}\\
    \multicolumn{1}{|c|}{} &  \multicolumn{1}{c|}{(2016)} &   \multicolumn{1}{c|}{} & \multicolumn{1}{c|}{(2016)} &   \multicolumn{1}{c|}{}\\

    \midrule
     A1 & 0.01  &  0.01  &  0.7500 & 0.7500 \\
     A7 & 0.03   &  0.01  & 0.6944 & 0.6944  \\
    A14 & 144.91  &  4.99  &  0.5333  &  0.5333 \\
    A15 &  0.54  &  0.17  &  0.6992 & 0.6992 \\
    A17 & 42.32  &  3.51  & 0.5773  & 0.5773 \\
    A20 & 14.55 &  0.11  & 0.7938  & 0.7938\\
    A21 & 305.48  & 15.08 & 0.5879 & 0.5879  \\
    A25 & 48743.90 & 678.53 &  0.5329  &   0.5329 \\
    A30 & 41.53 & 8.58  &  0.6304 &  0.6304 \\
   \bottomrule
    \end{tabular}  
  \label{table_inconsistent_results}%
\end{table}

Several instances provided by \citet{9}, which are different from its original sources  (papers from \citet{13} to \citet{35}, see Table~\ref{testset_a_problems}),  have been also used by \citet{62}.
These instances are A1, A7, A14, A15, A17, A20, A21, A25 and A30. For a fair comparison we have also run our model using the same input data (see Table ~\ref{table_inconsistent_results}).
Our experiments have confirmed all the results obtained by \citet{62}. Also we can conclude that the running times of our model have not changed much on these input data.

\subsection{Testet B results}

Since the test instances from Table~\ref{testset_b_problems} are less popular in research papers our goal is just to obtain optimal solutions for this set.
We have used our multi-start local search heuristic \citep{11} to get good solutions which are then passed as initial values for $\lambda$ parameter (we pick the best solution found by the heuristic within 30 seconds). 

The results for Testset B are shown in Table~\ref{testset_b_results}. Here we have found optimal solutions for 31 of 32 test problems.
Another result is an excellent performance of our multi-start local search heuristic algorithm: only one of 32 instances solved by the heuristic differs from the exact solution (instance B18). Due to the high computational complexity we are unable to solve the largest problem in the set -- problem B32 (50 $\times$ 150).

\section{Conclusion}
\label{sec_conclusion}
The cell formation problem is a well known combinatorial optimization problem with a  high computational complexity. 
A very few authors have suggested exact approaches for the most popular problem formulation with the grouping efficacy objective function. The majority of these works assume that the number of production cells is predefined. In this paper we suggest a new compact and effective integer linear programming model for the cell formation problem with a variable number of production cells. The model is based on the machine-machine and part-machine relations instead of the widely used machine-part-cell relation. It allows us to drastically reduce the number of variables and constraints in the resulting integer linear program.
Computational experiments show that our new model outperforms the state-of-art exact methods. We have solved 63 of 67 problem instances to the global optimum with respect to a variable number of production cells. We have also found several new solutions unknown before.
Unfortunately many problem instances from the cell formation datasets have inconsistencies with the original papers. This makes it really difficult to perform computational experiments and compare results to other approaches in the field. We have extracted and checked over 67 problem instances. All these data are available for downloading from website 
 \href{http://opt-hub.com/problems/cfp}{opt-hub.com} or \href{https://www.researchgate.net/publication/316648108_Test_instances_and_solutions_for_the_cell_formation_problem}{researchgate.net} and we hope it will help the researchers in this area. 
 The suggested model can be also used for solving biclustering problems and this is one of the directions of our future work.

\section{Acknowledgments}
This work was conducted at National Research University Higher School of Economics, Laboratory of
Algorithms and Technologies for Network Analysis and supported by RSF grant 14-41-00039.\newline
\label{sec_ack}



\begin{table*}[t]
 \centering
 \captionsetup{font=small}
 \setlength{\tabcolsep}{4pt}
  \normalsize
  \caption{Testset A - Computational results (residual cells are prohibited, singletons are allowed) }
    \begin{tabular}{|c|rrr|lrr|}
    \toprule
    \multicolumn{1}{|c|}{\multirow{4}{*}{\textbf{\#}}}  &  \multicolumn{3}{c|}{\textbf{Time, sec}} & \multicolumn{3}{c|}{\textbf{Efficacy}} \\
    \multicolumn{1}{|c|}{} &  \multicolumn{1}{c|}{Elbenani \& }& \multicolumn{1}{c|}{Bychkov}    & \multicolumn{1}{c|}{}    &\multicolumn{1}{c|}{Elbenani \&} & \multicolumn{1}{c|}{Bychkov} & \multicolumn{1}{c|}{} \\
    \multicolumn{1}{|c|}{} &  \multicolumn{1}{c|}{Ferland (2012)}&  \multicolumn{1}{c|}{et al.} & \multicolumn{1}{c|}{two-index} & \multicolumn{1}{c|}{Ferland (2012)} & \multicolumn{1}{c|}{et al.} &\multicolumn{1}{c|}{ two-index } \\
    \multicolumn{1}{|c|}{} &  \multicolumn{1}{c|}{\textbf{}}&  \multicolumn{1}{c|}{(2014)}  & \multicolumn{1}{c|}{model} & \multicolumn{1}{c|}{(cells)} & \multicolumn{1}{c|}{(2014)}  & \multicolumn{1}{c|}{model} \\
    \midrule
     A1 & 2.3 &0.63  &  0.00  & \qquad0.8235(2) & 0.8235 & 0.8235\\
     A2 & 1.6 &2.29  &  0.00   &\qquad0.6957(2) & 0.6957 & 0.6957  \\
     A3 & 3.1 &5.69  &  0.00  & \qquad0.7959(2) & 0.7959 &  0.7959   \\
     A4 & 2.0   &1.86  &  0.09   & \qquad0.7692(2) & 0.7692  &  0.7692  \\
     A5 & 30.6 &9.14  &  0.17   & \qquad0.6087(5) & 0.6087 & 0.6087 \\
     A6 & 4.3  &5.15   &  0.01  & \qquad0.7083(4) & 0.7083 & 0.7083 \\
     A7 & 9.6  &13.37   &  0.02  & \qquad0.6944(4) & 0.6944 & 0.6944 \\
     A8 & 3.1  &18.33   &  0.01  & \qquad0.8525(3) & 0.8525 & 0.8525\\
     A9 & 3.5 &208.36  &  0.45   &\qquad0.5872(2) & 0.5872 & 0.5872\\
    A10 & 1.1 & 6.25   &  0.00  & \qquad0.7500(5) & 0.7500  & 0.7500\\
    A11 & 1.6 & 2.93  &  0.02  & \qquad0.9200(3) & 0.9200  & 0.9200 \\
    A12 & 2188.7 & 259.19 & 0.19  & \qquad0.7206(7) & 0.7206  &  0.7206 \\
    A13 & 593.2 & 179.21 & 0.23   & \qquad0.7183(7) & 0.7183 &  0.7183 \\
    A14 & 15130.5 & *  &  4.24  & \qquad0.5326(8) & *  &  \textbf{0.5326}\\
    A15 & 252.5 & *  &  0.25  & \qquad0.6953(6)\textsuperscript{E} & * & \textbf{0.6899}\\
    A16 & 183232.5 & *  &  4.80  & \qquad0.5753(8) & *  & \textbf{0.5753}\\
    A17 & 2345.6 & *  &  3.82  & \qquad0.5773(9) & *  & \textbf{0.5773}\\
    A18 & * & *      & 32243.10 & \qquad* & * & \textbf{0.4345}  \\
    A19 & 131357.5 & *  & 245.59  & \qquad0.5081(7) & *  & \textbf{0.5081}\\
    A20 & 31.1 & *   &  0.22  & \qquad0.7791(5) & *  &  \textbf{0.7791} \\
    A21 & 14583.6 & *    &  24.34 & \qquad0.5798(5) & * &  \textbf{0.5798}  \\
    A22 & 11.3 &1.64  & 0.14  & \qquad1.0000(7) & 1.0000     & 1.0000 \\
    A23 & 230.7 & *   &  0.12   & \qquad0.8511(7) & *  & \textbf{0.8511} \\
    A24 & 1101.1 & *   &  0.16  & \qquad0.7351(7) & *  & \textbf{0.7351} \\
    A25 & * & *       &  1026.96 & \qquad* & *  &   \textbf{0.5329} \\
    A26 & * & *          & 178182.24  & \qquad* & *  & \textbf{0.4895} \\
    A27 & * & *           & *  & \qquad* & * & * \\
    A28 & 958714.1 & *    &  1964.00 & \qquad0.5482(5) & * &  \textbf{0.5482}\\
    A29 & * & *      & *  & \qquad* & *  & * \\
    A30 & 378300.0 & *     &  8.72  &\qquad0.6331(14) & *&  \textbf{0.6331} \\
    A31 & * & *     & 136.00 & \qquad0.6012(13)\textsuperscript{E} & * & \textbf{0.5977} \\
    A32 & * & *     & * & \qquad* & *  & *  \\
    A33 & * & *     & * & \qquad*  & * & \underline{0.4800} \\
    A34 & 268007.6 & *   & 16323.71 & \qquad0.6064(3) & *   & \textbf{0.6064}\\
    A35 & 7365.3 & *  &  1.34 & \qquad0.8403(10) &  * & \textbf{0.8403} \\
    \bottomrule
    \end{tabular}
  \label{testset_a_single}%
\end{table*}

\begin{table*}[t]
 \centering
 \captionsetup{font=small}
 \setlength{\tabcolsep}{4pt}
  \normalsize
  \caption{Testset A - Computational results (residual cells are allowed) }
    \begin{tabular}{|c|rrrr|llr|}
    \toprule
    \multicolumn{1}{|c|}{\multirow{4}{*}{\textbf{\#}}}  &  \multicolumn{4}{c|}{\textbf{Time, sec}} & \multicolumn{3}{c|}{\textbf{Efficacy}} \\
    \multicolumn{1}{|c|}{} &  \multicolumn{1}{c|}{}& \multicolumn{1}{c|}{Pinheiro et al.} & \multicolumn{1}{c|}{Pinheiro et al.}   & \multicolumn{1}{c|}{}    &\multicolumn{1}{c|}{Brusco} & \multicolumn{1}{c|}{Pinheiro} & \multicolumn{1}{c|}{} \\
    \multicolumn{1}{|c|}{} &  \multicolumn{1}{c|}{Brusco}&  \multicolumn{1}{c|}{(2016)} &  \multicolumn{1}{c|}{(2016)} & \multicolumn{1}{c|}{two-index} & \multicolumn{1}{c|}{(2015)} & \multicolumn{1}{c|}{et al.} &\multicolumn{1}{c|}{two-index} \\
    \multicolumn{1}{|c|}{} &  \multicolumn{1}{c|}{(2015)}&  \multicolumn{1}{c|}{IM} &  \multicolumn{1}{c|}{ILP} & \multicolumn{1}{c|}{} & \multicolumn{1}{c|}{(cells)} & \multicolumn{1}{c|}{(2016)}  & \multicolumn{1}{c|}{\textbf{}} \\
    \midrule
     A1 & 0.01 & 0.16 & 0.01  &  0.01  & \qquad0.8235(2,3,4) & 0.7500\textsuperscript{E} & 0.8235 \\
     A2 & 0.01 & 0.07 & 0.01  &  0.01   & \qquad0.6957(2,3,4) & 0.6956 & 0.6957  \\
     A3 & 0.02 & 0.09 & 0.03  &  0.01  & \qquad0.8085(2,3,4) & 0.8085 &  0.8085   \\
     A4 & 0.01 & 0.02  & 0.01  &  0.01   & \qquad0.7916(2,3,4) & 0.7917  &  0.7917  \\
     A5 & 0.6  & 0,29 & 0.06  &  0.17   & \qquad0.6087(2,3,4,5,6) & 0.6087 & 0.6087\\
     A6 & 0.04  & 0.14 & 0.01   &  0.01  & \qquad0.7083(2,3,4,5) & 0.7083 &  0.7083  \\
     A7 & 0.08  & 0.18 & 0.03   &  0.01  & \qquad0.6944(2,3,4,5) & 0.6944\textsuperscript{E} & 0.6944  \\
     A8 & 0.01  & 2.06 & 0.04   & 0.01  & \qquad0.8525(2,3,4) & 0.8525 & 0.8525 \\
     A9 & 35.86 & 81.46 & 4.94  &  0.45   & \qquad0.5872(2,3,4) & 0.5872 & 0.5872\\
    A10 & 0.06 & 0.03 &  0.01  &  0.01  & \qquad0.7500(2,3,4,5,6) & 0.7500  & 0.7500 \\
    A11 & 0.01 & 0.01 & 0.02  &  0.02  & \qquad0.9200(2,3,4) & 0.9200  & 0.9200 \\
    A12 & 633.91 & 0.49 & 0.09 & 0.03  & \qquad0.7424(6,7,8) & 0.7424  & 0.7424 \\
    A13 & 2631.76 & 0.49 & 0.11 & 0.03   & \qquad0.7285(6,7,8) & 0.7286 &  0.7286 \\
    A14 & 24716.34 & 600.98 & 144.91  &  4.88  & \qquad0.5385(8) & 0.5333\textsuperscript{E}  &  0.5385 \\
    A15 & 1279.93 & 7.24 & 0.54  &  0.16  & \qquad0.6992(5,6,7) & 0.6992\textsuperscript{E} & 0.6992 \\
    A16 & - & 1156.23  & 125.62 &  4.24  & \qquad- & 0.5804  & 0.5804  \\
    A17 & 20840.55 & 87.13 & 42.32  &  3.84  & \qquad0.5773(9) & 0.5773\textsuperscript{E}  & 0.5773 \\
    A18 & - & *   & *   & 52810.10 & \qquad- & * & \textbf{0.4397}  \\
    A19 & 1375608.66 & 23928.70& 1771.99  & 249.52  & \qquad0.5081(7) & 0.5081  & 0.5081 \\
    A20 & 4830.00 & 1.78  & 14.55 &  0.09  & \qquad0.7888(5,6,7) & 0.7938\textsuperscript{E}  & 0.7888\\
    A21 & - & 2145.24  & 305.48  & 22.60 & \qquad- &0.5879\textsuperscript{E} & 0.5860  \\
    A22 & 0.01 & 0.02 & 0.15  &  0.14  & \qquad1.0000(7) & 1.0000     & 1.0000\\
    A23 & 42.29& 10.08 & 0.44   &  0.14   & \qquad0.8511(7) & 0.8511  &  0.8511  \\
    A24 & 208158.02 & 17.46 & 0.78   &  0.20 & \qquad0.7351(7) & 0.7351  & 0.7351 \\
    A25 & - & 371233.00  &  48743.90 & 759.70 & \qquad- & 0.5329\textsuperscript{E}  &   0.5329 \\
    A26 & - & *  & * & 134418.65 & \qquad- & *  & \textbf{0.4895} \\
    A27 & - & *  & * & * & \qquad- & * & * \\
    A28 & - & *  &  * & 46361.97 & \qquad- & * & \textbf{0.5482} \\
    A29 & - & *  & * & * & \qquad- & *  & * \\
    A30 & - & 183.71  &  41.53 & 8.00  & \qquad- & 0.6304\textsuperscript{E} &  0.6331 \\
    A31 & - & 13807.50  & 2622.06 & 64.82 & \qquad- & 0.5977 & 0.5977  \\
    A32 & - & *  & * & 234055.90 & \qquad- & *  & \textbf{0.5084} \\
    A33 & - & *  & * & * & \qquad-  & * & \underline{0.4829} \\
    A34 & - & *  & * & 14212.57 & \qquad- & *   & \textbf{0.6131} \\
    A35 & - & 325.53  &  18.22 & 1.61 & \qquad- &  0.8403 & 0.8403 \\
    \bottomrule
    \end{tabular}
  \label{testset_a_residual}%
\end{table*}

\begin{table*}[t]
  \centering
  \normalsize
 \setlength{\tabcolsep}{6pt}
 
  \caption{Testset B - Computational results}
    \begin{tabular}{|crrrrr|}
    \toprule
    \multicolumn{1}{|c}{\multirow{4}[4]{*}{\textbf{\#}}} & \multicolumn{2}{c}{\textbf{Time}} & \multicolumn{3}{|c|}{\textbf{Efficacy}}  \\
    \multicolumn{1}{|c}{} & \multicolumn{1}{|c|}{two-index}  & \multicolumn{1}{c}{two-index} & \multicolumn{1}{|c|}{Heuristic}  & \multicolumn{1}{c|}{two-index}  & \multicolumn{1}{c|}{two-index} \\
    \multicolumn{1}{|c}{} & \multicolumn{1}{|c|}{(no residual} & \multicolumn{1}{c}{(allow } & \multicolumn{1}{|c|}{bound} &  \multicolumn{1}{c|}{(no residual} & \multicolumn{1}{c|}{(allow } \\
 \multicolumn{1}{|c}{}  & \multicolumn{1}{|c|}{cells)}& \multicolumn{1}{c}{residual cells)} & \multicolumn{1}{|c|}{} & \multicolumn{1}{c|}{cells)}& \multicolumn{1}{c|}{residual cells)}\\
   \midrule
  B1  & 0.01 & 0.01  &0.8095  & 0.8095 & 0.8095\\
  B2  & 0.01 & 0.01  &0.7222  &0.7222& 0.7222\\
  B3  & 0.25 & 0.03  &0.6071  & 0.6071 & 0.6071\\
  B4 & 0.01 & 0.01  &0.8889  & 0.8889 & 0.8889\\
  B5 & 0.01 & 0.01 &0.7500  & 0.7500 & 0.7500 \\
  B6 & 0.01 & 0.01  &0.7391  & 0.7391 & 0.7391 \\
  B7 & 0.01 & 0.01  &0.8148  & 0.8148 & 0.8148 \\
  B8 & 0.01  & 0.01  & 0.7222 & 0.7222 & 0.7222 \\
  B9 & 0.01 & 0.01  & 0.7576  & 0.7576& 0.7576  \\
  B10 & 0.01 & 0.01  & 0.9000 &0.9000 & 0.9000 \\
  B11 &  0.01 & 0.02  & 0.7273 & 0.7273 & 0.7297 \\
  B12 & 0.01 & 0.01  & 0.8276  &0.8276 & 0.8276 \\
  B13 & 0.36 &0.80  & 0.5962  & 0.5962 & 0.6042 \\
  B14 & 0.25  &0.30  & 0.6405  & 0.6405& 0.6405 \\
  B15 & 0.01  &0.01  & 0.8333  & 0.8333& 0.8333 \\
  B16 & 0.16  &0.06  & 0.7391  & 0.7391  & 0.7444\\
  B17 & 0.98 & 0.26  & 0.6552  & 0.6552  & 0.6842\\
  B18  & 1.82 & 1.65  &0.6027 & 0.6129 & 0.6129\\
  B19 & 0.03 & 0.06  & 0.8000  & 0.8000  & 0.8113\\
  B20 & 0.05 & 0.03  &0.8710  & 0.8710  & 0.8710\\
  B21 & 0.03 & 0.04  &0.8333  & 0.8333  & 0.8333\\
  B22 & 0.05 & 0.01  &0.7258  & 0.7258 & 0.7258 \\
  B23 & 0.05  & 0.06  &0.8111  & 0.8111  & 0.8111\\
  B24 & 4.79 & 7.80  & 0.5673  & 0.5673   & 0.5728\\
  B25 & 0.20 & 0.10  &0.7600  & 0.7600  & 0.8000\\
  B26 & 13.81 & 25.75  &0.6068  & 0.6068 & 0.6078 \\
  B27 & 0.25 & 0.28  &0.7248  & 0.7248  & 0.7248\\
  B28 & 0.83 & 1.04  &0.6729  & 0.6729  & 0.6729\\
  B29 & 33.82 & 51.76  &0.5730  & 0.5730  &0.5745\\
  B30 & 4.76 & 8.67  &0.7308  & 0.7308  & 0.7325\\
  B31 & 19.69 & 17.50  &0.6799  & 0.6799  & 0.6799\\
  B32 & * & *  &0.6193  & *   & * \\ 
    \bottomrule
    \end{tabular}%
  \label{testset_b_results}%
\end{table*}%

\end{document}